\documentclass[12pt]{iopart}

\usepackage{graphicx}
\usepackage{psfrag}

\newcommand\half{{\textstyle{\frac{1}{2}}}}

\newcommand{\Ec}{{\cal E}_{\rm C}}

\begin{document}

\title[The Casimir effect as scattering problem]{The 
Casimir effect as scattering problem}

\author{A Wirzba}
\address{Institut f\"ur Kernphysik,
         Forschungszentrum J{\"u}lich,
         D-52425 J{\"u}lich, Germany}
\ead{a.wirzba@fz-juelich.de}

\begin{abstract}
We show
that Casimir-force calculations for a finite number of
non-overlapping obstacles can be mapped onto 
quantum-mechanical  
billiard-type problems  which are characterized by
the scattering of a fictitious
point particle off the very same obstacles.
With the help of a modified
Krein trace formula the genuine/finite part of the Casimir energy
is determined as
the energy-weighted integral over
the log-determinant of the multi-scattering matrix of the analog
billiard problem.  The formalism is self-regulating and inherently 
shows that
the Casimir energy is governed by the infrared end of
the multi-scattering phase shifts or spectrum of the fluctuating field.
The calculation is exact and in principle applicable for any
separation(s) between the obstacles. In practice, it is more suited
for large- to medium-range separations.
We report especially about the Casimir energy of a fluctuating massless
scalar field between two spheres or a sphere and
a plate under Dirichlet and Neumann boundary conditions. But
the formalism can   easily be
extended to any number of spheres and/or planes in
three or arbitrary dimensions, with a variety of boundary conditions or
non-overlapping potentials/non-ideal reflectors.

\end{abstract}



\section{Introduction}

In 1948, the Dutch physicist H~B~G~Casimir predicted the remarkable effect
\cite{casimir48} that
two parallel, very closely spaced, uncharged metallic
plates 
attract each other  in vacuum. 
The origin of this force can be traced back to the modifications of the
zero-point fluctuations of the electromagnetic field by the presence of
the two plates (at distance $L$) relative to the free case, or rather, 
relative to the case where
the plates are separated by an infinite distance. Mathematically, 
this corresponds to the following difference of two mode sums:
\[
  \left.\sum \half \hbar \omega_k\right|_{{\rm plates}(L)}
 -\left.\sum \half \hbar \omega_k\right|_{{\rm plates}(L\to\infty)}\ .
\]
The distinctive feature of the Casimir effect is that it depends on
the geometry in a non-intuitive way: its strength and, 
perhaps, its
sign are geometry-dependent (for a review see~\cite{bordag}).

Let us look at the other side of the `coin' and take the geometry dependence
as guiding principle for the construction of the Casimir effect; 
{\it i.e.} we will
define the Casimir energy as that part of the vacuum energy that results from
the geometry-dependent and therefore changeable part of the pertinent 
density of states. In fact, in Casimir-type calculations, the density of states
$\rho(E)=\sum_{E_k}\delta(E-E_k)$, where $\{E_k\}$ 
are the eigenenergies of the modes, can be split into three parts:
\begin{equation}
\rho(E) = \rho_0(E)
+ \rho_{\rm bulk}(E) +\delta\rho_{{\rm C}}(E)\,.
\end{equation}
(i) There is  
a free part $\rho_0(E)$ over the homogeneous part of the background (the
fluctuating field, matter fields etc.)
that is completely unaffected by the presence of the obstacles; (ii) the bulk 
part $\rho_{\rm bulk}(E)$
sums up the  modifications of the spectrum 
(excluded volumes and surface
contributions including Friedel oscillations)  
by each of the individual obstacles  as though they were still 
infinitely separated; (iii) finally the remaining part  
$\delta\rho_{\rm C}(E)$ is the {\em genuinely geometry-dependent} 
part of the
density of states which `knows' about the relative separations and angles
between the various obstacles. It is the latter that determines the
Casimir energy when it is integrated up, weighted by the energy $E$:
\begin{equation}
 \Ec  \equiv \half\int_0^\infty \rmd E \, E\, \delta\rho_{\rm C}(E)
= - \half\int_0^\infty \rmd E\, {\cal N}_{\rm C}(E)  \,, \label{Ecgeo}
\end{equation}
where 
$
{\cal N}_{\rm C}(E)=\int_0^E \rmd E'\,\delta\rho_{\rm C}(E')
$
is the geometry-dependent part of the 
integrated density of states 
(or number of states),
\begin{equation}
{\cal N}(E) \equiv \int_0^E \rmd E'\,\rho(E')= {\sum_{E_k}} \Theta(E-E_k)
\,.
\end{equation}
Note that the integrals in (\ref{Ecgeo}) do not converge absolutely, but 
only conditionally because of the oscillating behaviour of 
$\delta\rho_{\rm C}(E)$
and ${\cal N}_{\rm C}(E)$. The inclusion 
of {\it e.g.} an exponential 
damping factor (compare with the analogous tilt of the integration path(s)
in (\ref{ECresult})),  which
can be removed after the integration(s) are performed,  
cures this problem and also eliminates
the spurious upper-boundary contribution 
$\half E {\cal N}_{\rm C}(E)|^\infty$ 
from the  partial integration. The lower boundary term  
$\half E {\cal N}_{\rm C}(E)|_0$  vanishes automatically.

\section{The mapping of the Casimir calculation  
onto a scattering problem}

The Casimir calculation for two parallel plates at distance $L$ was simple
since the problem is separable, {\it i.e.} quasi-one-dimensional  
if the plates had infinite extent.
For more complicated geometries than two parallel plates
only the proximity
force approximation~\cite{proximity} is left in general (see, however,
section~\ref{sec:fin_rem}).

However, as shown in \cite{BW01,BMW05} for the non-relativistic 
fermionic Casimir effect and
in \cite{BMW06,WBM06} 
for the scalar Casimir effect\footnote{In the scalar 
Casimir
effect the fluctuating
field is not the electromagnetic one, but a massless scalar one with
the dispersion $E=\hbar c|\vec k|$. In the 
non-relativistic
fermionic Casimir effect, the background does not consist 
of fluctuating fields, 
but of (a Fermi sea of) non-relativistic fermionic matter waves.}, 
the Casimir energy between 
any 
finite number of non-overlapping {\em spherical} obstacles, {\em i.e.} 
spheres and
cylinders in 3
dimensions or disks in 2 dimensions, can be solved exactly, although the
corresponding 
quantum-mechanical problems are not separable any longer. In fact, these
calculations simplify because
of Krein's trace formula~\cite{krein}
which maps -- under the condition that the potentials of the obstacles
are sufficiently short-ranged -- 
the difference in the total level 
densities -- with and without obstacles -- to
the energy variation of the total phase 
shift\footnote{See  \cite{uhlenbeck} 
for a precursor of Krein's formula for potentials with spherical symmetry.}.
For the special  case of $n$ spherically symmetric obstacles of radii $a_i$ at 
mutual center-to-center separation
$r_{ij}> a_i + a_j$ (such that they do not overlap) and mutual angles 
$\alpha_{ij}$ the Krein formula reads
\begin{eqnarray}
\delta\bar\rho(E) 
= \bar\rho(E)-\bar\rho_0(E)
&=& \frac{1}{2\pi \rmi} \;\frac{\rmd\, }
{\rmd E}\tr \ln S_n \bigl(E,\{a_i\},\{\vec r_{ij}\}\bigr)\nonumber\\
&=& \frac{1}{2\pi \rmi} \;\frac{\rmd\, }
{\rmd E}\ln \det S_n \bigl(E,\{a_i\},\{\vec r_{ij}\}\bigr) 
\,,
\label{Krein}
\end{eqnarray}
where the distances $r_{ij}$ and angles $\alpha_{ij}$ have been 
combined to separation vectors $\vec{r}_{ij}$ and where $i,j=1, 2,\cdots, n$
label the obstacles. 
Note that the level densities on the left hand side are  averaged
over an energy-interval larger than the mean-level spacing in the volume
$V$ of the entire system which will be taken to infinity in the end. This
averaging is done 
in order to match to the right hand side which is
defined in terms of the total $S$-matrix 
$S_n\bigl(E,\{a_i\},\{\vec r_{ij}\}\bigr)$
of the $n$-disk system in 2 dimensions~\cite{gaspard,wreport,wh98} or
the $n$-sphere system in 3 dimensions~\cite{hwg97}, respectively, 
and which is therefore continuous.
In fact, in \cite{wreport,wh98} it was shown that the 
determinant of $n$-disk
S-matrix is finite (except at the known poles and branch points), although
the matrix has an infinite number of components,
since the pertinent T-matrix was proved to be trace class. The 3-dimensional
$n$-sphere case was discussed in \cite{hwg97}.
In this way
the Casimir calculation for a finite number of non-overlapping 
obstacles is mapped onto the  quantum-mechanical analog of a
classical {\em billiard-type} problem 
which is characterized by the scattering of
a fictitious point particle off the very same obstacles. For the
case of only one spherical
obstacle the scattering problem is separable, but uninteresting as there is
no Casimir effect. For two 
spherical obstacles the scattering problem is non-separable and classically 
hyperbolic, and for more than
two obstacles it is in general even classically 
chaotic~\cite{eckhardt,gaspard,cvitanovic,W92,W93}.

Let us remark that the energy-averaging and the infinite volume limit do 
not commute,
but that first the volume has to be taken to infinity and only then the
averaging interval can go to zero. This is obvious if one takes 
semiclassical considerations into account
where the determinant over the $n$-sphere/disk matrix is given in 
terms of periodic orbits and possible Weyl term corrections, see 
\cite{wreport,wh98}. For this purpose let us put the scattering system
at the middle of a large container which can be 
taken, without loss of generality, to
be spherical of finite radius $R$ 
and consider this container together with its empty reference
container that is of equal size.
Now, as long as the containers have a finite size, there exist three classes
of periodic orbits. (i) The generic scattering periodic orbits that only
bounce between the obstacles belonging to the scattering system and that 
semiclassically specify the determinant of the scattering matrix.
(ii) A first class of spurious periodic orbits which solely are there 
because of the container and which only bounce between the walls of the
container. The subtraction of the reference system, in fact, removes this
class of spurious orbits. (iii) A second class  of spurious periodic orbits
which bounce between the container walls {\em and} the scattering obstacles
and which do not exist in the empty reference container.
As the length $L_{\rm p.o.}$ 
of these periodic orbits will increase with the size of the
container, their contribution to the periodic orbit sum  
can be suppressed if a small imaginary term $\rmi\epsilon$
is added to the energy or wave number $k\equiv |\vec k|$ which is here
the same because  the dispersion is $E=\hbar c |\vec k|$. 
The phase of the orbit
acquires therefore an exponential damping factor
\begin{eqnarray}
       \rme^{\rmi k L_{\rm p.o}} \to \rme^{\rmi(k+\rmi\epsilon) L_{\rm p.o}}
\end{eqnarray} 
which smooths (averages) 
the total and background densities of states in the Krein
formula 
\begin{eqnarray*}
\lim_{\epsilon\to 0_+}\lim_{R\to \infty}\left\{
\rho(k\!+\!\rmi\epsilon,R)-\rho_0(k\!+\!\rmi\epsilon,R)\right\}
= \frac{1}{2\pi \rmi}  \frac{\rmd\ }{\rmd E}
\tr \ln S_n \bigl(E(k),\{a_i\},\{\vec r_{ij}\}\bigr).
\end{eqnarray*}
This explains the smoothing/averaging procedure in (\ref{Krein}).

\section{The mapping to the multi-scattering matrix}

In the following, we will show that the  geometry-dependent Casimir 
fluctuations can be extracted 
from the {\em multiple}-scattering part
of the S-matrix. For this purpose let us once more explain the philosophy:
(i)  we start out with a large, but finite 
container of radius $R$ which will be filled with 
the background field (here the massless scalar field 
or a fermionic matter field) and which eventually
will be taken to infinite size. The total density of states is therefore
equal to the background density:
$  \rho(k+\rmi\epsilon,R)= \rho_0(k+\rmi\epsilon,R)$.
A suppression term is added to the wave number $k$ as explained in the last
section.
(ii)~Next we will also put the spherical obstacles of radii $a_i$ 
into the large container, 
where we make sure that the mutual separation $r_{ij}$ between any pair of 
obstacles is large, {\it i.e.} comparable with the size of the container.
The total density of states has to be readjusted because the background field
is affected by excluded volumes due to the obstacles  and  by 
boundary effects (including Friedel oscillations) due to the
imposed boundary conditions at the surface of the obstacles, say Dirichlet
boundary conditions:
\begin{equation}
  \rho(k+\rmi\epsilon,R,\{a_i\})= \rho_0(k+\rmi\epsilon,R)+\sum_{i=1}^n 
\underbrace{\rho_{\rm W}(k+\rmi\epsilon,R,a_i)}_{\mbox{Weyl term}}\,.
\end{equation}
(iii) Now, the obstacles are moved close to each other such that
the actual 
system of a finite number of non-overlapping spherical
obstacles  is formed. Again the density of states has to be readjusted, this
time by the geometry-dependent term $\delta \rho_{\rm C}$ which depends on
all radii $\{a_i\}$ and on all separation vectors $\{\vec r_{ij}\}$:
\begin{eqnarray}
  \rho(k+\rmi\epsilon,R,\{a_i\},\{\vec r_{ij}\})
  &=& \rho_0(k+\rmi\epsilon,R)+\sum_{i=1}^n 
\rho_{\rm W}(k+\rmi\epsilon,R,a_i)\nonumber\\
 &&\mbox{}
+\delta\rho_{\rm C}(k+\rmi\epsilon,R,\{a_i\},\{\vec r_{ij}\})\,.
\label{rho0WC}
\end{eqnarray}
(iv) By inserting the difference of (\ref{rho0WC}) and the background 
density of states $\rho_0$ into the Krein formula and by 
taking the double limit $\lim_{\epsilon\to0_+}\lim_{R\to\infty}$ we
effectively 
remove the container walls and map the problem to the scattering system:
\begin{eqnarray}
&&\lim_{\epsilon\to0_+}\lim_{R\to\infty}
\Bigl\{\rho\bigl((k+\rmi\epsilon,R,\{a_i\},\{\vec r_{ij}\}\bigr)
-\rho_0(k+\rmi\epsilon,R)\Bigr\} \nonumber\\
&&\equiv\bar\rho\bigl(k,\{a_i\},\{\vec r_{ij}\}\bigr)-\bar\rho_0(k) 
= \frac{1}{2\pi \rmi}\frac{\rmd\ }{\rmd k}
\ln\det S_n\bigl(k,\{a_i\},\{\vec r_{ij}\}\bigr)\,,
\end{eqnarray}
where $\ln\det S_n\bigl(k,\{a_i\},\{\vec r_{ij}\}\bigr)$ 
is the total phase shift
expressed through the $n$-sphere/disk S-matrix 
$S_n\bigl(k,\{a_i\},\{\vec r_{ij}\}\bigr)$.
As shown in \cite{wreport,wh98} and \cite{hwg97}, respectively, 
the determinant of the $n$-disk/sphere S-matrix
separates into a product of the determinants of the 1-disk/sphere
S-matrices $S_1(E,a_i)$, where $a_i$ is the radius of the single scatterer $i$,
and the ratio of the determinant of the inverse 
multi-scattering matrix $M$ and its complex conjugate:
\begin{equation}
\det{S_n\bigl( k,\{a_i\},\{\vec r_{ij}\} \bigr)} =
\left\{\prod_{i=1}^n \det{S_1(k,a_i)}\right\}
\frac{{\det \left(M(k^*,\{a_i\},\{\vec r_{ij}\})^\dagger\right)}}
{\det{ M}\left(k,\{a_i\},\{\vec r_{ij}\}\right)}
\label{split}
\, .
\end{equation}
Note that each and every determinant in (\ref{split}) is finite and
exists separately, since the pertinent T-matrices are  all
trace class~\cite{wreport,wh98,hwg97}.  

When inserted into Krein's formula,
the product over the single-scatterer determinants generates
just the bulk (or Weyl term) contribution  to the density of states
\begin{eqnarray}
 &&\lim_{\epsilon\to 0_+}\lim_{R\to\infty} 
   \rho_{\rm bulk}(k+\rmi\epsilon,R,\{a_i\})
  \equiv \sum_{i=1}^n  \lim_{\epsilon\to 0_+}\lim_{R\to\infty} 
\rho_{\rm W}(k+\rmi\epsilon,R,a_i)\nonumber\\
&& \equiv\sum_{i=1}^n  
\bar\rho_{\rm W}(k,a_i)
= \frac{1}{2\pi \rmi}\,\frac{\rmd\ }{\rmd k}\,
\sum_{i=1}^n \ln \det S_1(k,a_i)\,,
\end{eqnarray}
which takes care of the excluded volume terms and the surface terms (including
Friedel oscillations).
The geometry-dependent part of the density of states 
\begin{equation}
 \delta\bar\rho_{\rm C}(k,\{a_i\},\{ {\vec r}_{ij}\})
\equiv\lim_{\epsilon\to 0_+}\lim_{R\to\infty} 
\delta\rho_{\rm C}(k+\rmi\epsilon,R,\{a_i\},\{{ {\vec r}_{ij}}\})
\end{equation}
is therefore given by a modified
Krein equation~\cite{BW01} which is formulated in terms of the inverse
multi-scattering matrix 
{$M\left( k,\{a_i\},\{{ {\vec r}_{ij}}\}\right)$} 
instead of the full S-matrix
\begin{eqnarray}
{\delta\bar\rho_{\rm C}}(k,\{a_i\},\{ {\vec r}_{ij}\}) 
&=& \bar\rho(k,\{a_i\},\{ {\vec r}_{ij}\}) -\bar\rho_0(k)
-  \sum_{i=1}^n \bar\rho_{\rm Weyl}(k,a_i)\nonumber\\
&=& -{\frac{1}{\pi }}
{\rm Im}\,
 {\frac{\rmd}{\rmd k}}
 \ln \det { M(k,\{a_i\},\{\vec r_{ij}\})}\,.
 \label{modKrein}
\end{eqnarray}
Note that in this expression all the possibly divergent terms connected 
with the sharp-surface limits are subtracted out. In fact, by mapping
the system to a multi-scattering problem the calculation 
is self-regulating: an additional regulator is not needed (except
the energy-damping mentioned below (\ref{Ecgeo}) which
can be removed in the end, see also \cite{BMW06}). 

The pertinent Casimir energy can then be read off from the energy-weighted 
(or wave-number-weighted) integral 
\begin{eqnarray}
  \Ec &=& \int_0^\infty \rmd E\, \half E\,\delta\bar\rho_{ C}(E,\{a_i\},
  \{\vec r_{ij}\})
=  \half\hbar c \int_0^\infty \rmd k\, k \,\delta\bar\rho_{ C}(k,\{a_i\},
  \{\vec r_{ij}\})\nonumber\\
&=& -\half \hbar c\int_0^\infty \rmd k\, {\overline{\cal N}}_{\rm C}(k,\{a_i\},
  \{\vec r_{ij}\})=\frac{\hbar c}{2\pi}\int_0^\infty \rmd k\, 
{\rm Im} \ln\det M(k)\nonumber\\ 
&=& \frac{\hbar c}{4\pi \rmi}
\left[\int_0^{\infty(1+\rmi 0_+)}\rmd k \ln  \det M(k)-
\int_0^{\infty(1-\rmi 0_+)}\rmd k \ln  \det M(k^\ast)^\dagger\right]\nonumber\\
&=& \frac{\hbar c}{2\pi}\int_0^{\infty}\rmd k_4 \ln  \det M(\rmi k_4) \,,
\label{ECresult}
\end{eqnarray}
where the dependences on the radii and separation vectors were suppressed
in the argument of the
inverse multi-scattering matrix $M(k)\equiv M(k,\{a_i\},\{\vec r_{ij}\})$ for
simplicity. 
In the last step a Wick rotation $k\to \rmi k_4$ was performed in the first,
and $k\to -\rmi k_4$ in the second integral, and the relation
$\det M(\rmi k_4) = \det M(\rmi k_4)^\dagger$ was applied which follows 
from the relation 
$\det M(k)=\det M\left((-k)^\ast\right)^\dagger$~\cite{BMW06}.
As a corollary note that 
\begin{eqnarray}
&&\frac{\hbar c}{2\pi}\int_0^\infty\!\! \rmd k\, k^{2n+1}{\rm Im} \ln \det M(k)
\nonumber\\
&&=\rmi(-1)^n \frac{\hbar c}{4\pi}\int_0^\infty \!\!\rmd k_4\, k_4^{2n+1}
\left[\ln \det M(\rmi k_4)-\ln\det M(\rmi k_4)^\dagger\right] = 0\,,
\end{eqnarray}
{\it e.g.} the Casimir energy over modes with a non-relativistic dispersion
$E=\hbar^2 k^2/2m_{\rm N}$, where $m_{\rm N}$ is the pertinent mass, 
integrates to zero~\cite{BMW06}, unless there is a
{\em finite} upper cutoff, as {\it e.g.} the Fermi momentum $k_{\rm F}$ in
the case of the so-called
fermionic
Casimir effect~\cite{BW01}.

Note that the final expression (\ref{ECresult}) for the Casimir energy is
obviously finite, since $\det M(\rmi k_4)$ vanishes rapidly with increasing
value of $k_4$, and moreover, for the same reason, 
it is dominated by the infrared end of the
integration.

\section{The multi-scattering matrix and the fermionic Casimir effect}
\label{sec:fermionic}

The modified Krein equation (\ref{modKrein}) is especially useful as there
there exists a close-form expression for the inverse multi-scattering matrix
for $n$ spheres (of radii $a_j$ and mutual distances 
$r_{jj'}$, where the  indices $j,j'=1,2,\cdots,n$ label the spheres)
in terms of spherical
Bessel and Hankel functions of first kind, spherical harmonics and 
3j-symbols~\cite{hwg97}:
\begin{eqnarray}
&& M^{jj'}_{lm,l'm'} =\delta^{jj'}\delta_{ll'}\delta_{mm'}
                 + (1-\delta^{jj'})\,
\rmi^{2m+l'-l}\,\sqrt{4\pi(2l\!+\!1)(2l'\!+\!1)}\nonumber\\
&&\qquad\mbox{}\times
\left({\frac{a_j}{a_{j'}}} \right)^2
{\frac{{ j_l(k a_j)}}{{ h_{l'}^{(1)}(k a_{j'})}}}
\sum_{l'' =0}^{\infty}\sum_{m''=-l'}^{l'}
\sqrt{{2l''}+1}
 \,\rmi^{l''}\,
{\left({\begin{array}{ccc}
  {l''}& {l'}&{l} \\
  {0}&{0}&{0} \end{array}}\right)}\nonumber\\
&&\qquad\mbox{}\times
 {\left({\begin{array}{crc}
  {l''} & {l'} & {l}\\
  {m\!-\!m''} & {m''} & {-m} \end{array}}\right)}
D^{l'}_{m'\!,m''}(j,j')\,
{  h_{l''}^{(1)}(k r_{jj'})}\,
{ Y_{l''}^{m\!-\!m''}\bigl( \hat r^{(j)}_{jj'}\bigr)}.
\label{Msphere}
\end{eqnarray}
Here $l,l'$ and $m,m'$ are total angular momentum and associated magnetic
quantum numbers, respectively.
The rotation matrix $D^{l'}_{m'\!,m''}(j,j')$ maps the
local coordinate system of the sphere $j'$   to the one of the sphere
$j$ and the unit vectors $\hat r^{(j)}_{jj'}$  
point from the origin of the sphere $j$ (as measured in its local
coordinate system) to the origin of the sphere $j'$.
The above expression refers to Dirichlet boundary conditions on the
spheres. The case of Neumann boundary conditions (or 
even mixed boundary conditions) on the spheres follows from the replacement
\begin{equation}
  \frac{j_l(k a_j)}{h^{(1)}_{l'}(k a_j)}\longrightarrow
    \frac{ \frac{\rmd\ }{\rmd a}\left.\left(a j_l(k a)\right)\right|_{a=a_j}}
       { \frac{\rmd\ }{\rmd a'}\left.\left(a' h^{(1)}_{l'}(k a')\right)
\right|_{a'=a_{j'}}}
\end{equation} 
etc.
A  closed-form expression for the two-dimensional 
case of  $n$ disks, corresponding to (\ref{Msphere}),
can be found in \cite{wreport}.

For the case of small scatterers one can even simplify 
the expression of the multi-scattering matrix:
\begin{equation}
M^{jj'}(k,\{a_i\},\{r_{ij}\})\approx
\delta ^{jj'} - \bigl(1-\delta ^{jj'}\bigr)
{{f^{s}(k,a_j)}}  
 \frac{ \rme^{\rmi k{r_{jj'}}}}{{r_{jj'}}}
 +  {\cal O}(\mbox{$p$-wave}),
\end{equation}
since the $s$-wave scattering dominates over all other partial waves.
Thus,  the non-trivial part of the inverse multi-scattering matrix is given
by the propagation of spherical waves modulated by  $s$-wave amplitudes 
$f^{s}(k,a_j)$ between
the spheres. From the modified Krein formula 
the integrated density of states in the
case of two small spherical cavities of common radius $a$ and center-to-center
separation $r$ can be deduced as~\cite{BW01}  
\begin{equation}
{\cal{N}}_{\rm C}^{\rm oo}(k)=-\frac{1}{\pi}{\rm Im}\ln\det
M^{{\rm oo}}(k,a,r)\approx 
\frac{{ a^2}}{\pi { r^2}} \sin [ 2({ r}-{ a})k]
+ {\cal{O}}\left((ka)^3\right).
\label{s-wave}
\end{equation}
This expression should be compared with the semiclassical approximation
that sums up all partial waves
\begin{equation}
{\cal{N}}_{\rm C,\,sc}^{\rm oo}(k)= 
\frac{a^2}{4\pi r(r-2a)} \sin [ {{2(r-2a)}k}].
 \label{leadsc}
\end{equation}
The latter is  the leading contribution to Gutzwiller's trace
formula~\cite{gutbook}, namely the contribution of the non-repeated
two-bounce periodic orbit
between the two spheres. 
$S_{\rm po}(k)/\hbar = 2(r-2a)k$ is the action of the two-bounce 
periodic orbit,  where $2(r-2a)$ is the length of its
geometric path. Note that the
semiclassical result is suppressed by a
factor of $1/2^2$ {relative} to the small-scatterer one. We will see that
each factor $1/2$ is associated with a semiclassical reflection from one
sphere.  

It was shown in \cite{BW01}, under the condition $k> 1/a$, that 
the semiclassical 
result for ${\cal{N}}_{\rm C,\,sc}^{\rm oo}(k)$ is a very good
approximation of the full quantum-mechanical
result calculated from the exact expression
(\ref{Msphere}) of the two-sphere scattering matrix when plugged into
the modified Krein formula (\ref{modKrein}).
Therefore, the Casimir energy for two spherical cavities inside a Fermi sea
of non-relativistic non-interacting matter modes can be approximated
in terms of a spherical Bessel function $j_1$ as
\begin{eqnarray}
\Ec^{\rm oo}\ =\ - \int_0^{k_{\rm F}} \rmd k\, {{\cal N}_{\rm C}^{oo}(k,a,r)}
\approx -\frac{k_{\rm F}^2}{2 m_{\rm N}} 
\frac{a^2}{2\pi r(r-2a)} j_1[2(r-2a)k_{\rm F}]\,
\label{Ecf00}
\end{eqnarray}
where the expression is 
valid for $k_{\rm F} a >1$. Here $m_{\rm N}$ is the mass of the
fermionic mode. 
Note that it is long-ranged, {\it i.e.} it scales as $a^2/L^{3}$ 
with $L=r-2a$. 
The corresponding fermionic Casimir energy of
the sphere-plate system
\begin{eqnarray}
\Ec^{{\rm o} {\mathbf \vert}}
\approx - \frac{k_{\rm F}^2}{2m_{\rm N}} 
\frac{a}{2\pi (r-a)} j_1[2(r-a)k_{\rm F}]\,,
\label{Ecf0p}
\end{eqnarray}
even scales  as $a/L^{2}$ with $L= r-a$.
Moreover, the fermionic Casimir energy has in both cases an oscillating
behaviour: with increasing distance between the obstacles,
the Casimir energy, which starts out to be attractive, will become repulsive,
and under a further increase of the distance, it will become
attractive again, where the strength is of course reduced. 
This is in contrast to  the  fixed (negative) sign of 
the standard Casimir effect with
fluctuating electromagnetic or scalar fields between these obstacles. The
reason for this oscillating pattern in the fermionic Casimir effect is
the presence of a new scale, in addition to the length scale(s), namely the
Fermi momentum $k_{\rm F}$.
In fact,
the strength of this fermionic Casimir energy is determined by the
UV-cutoff of the theory~\cite{BW01}. Also
this behaviour distinguishes the fermionic Casimir effect from the standard
Casimir effect: the latter is governed by the infrared behaviour of the
corresponding density of states.

In \cite{BW01},  the fermionic Casimir energy of the three- and 
four-sphere system was calculated as well. From 
the periodic orbit summation it is
obvious that there exist genuine three- and more-body  interactions. However,
it was shown in \cite{BW01} that the two-bounce orbit dominates in
the equilateral three- and four-sphere systems. In fact, 
the billiard analogy holds: it is difficult to make {\em long} shots, 
especially with many bounces -- the slightest error ruins the shot.
Mathematically, the relative weights calculated from the instabilities  
of the longer periodic orbits, even
weighted with their degeneracy factors, are far less than the weight of
the two-bounce orbit. The maximal correction due to the next-to-leading
periodic orbit, the triangular three-bounce orbit, is at most 10\,\% at
a distance $r\approx 2.5 a$ and even smaller for other separations.  

\section{The sphere-plate case of the scalar Casimir effect}

{}From now on we will only discuss the scalar Casimir effect, where the 
fluctuating modes are massless scalar ones. This 
effect was first investigated by Schaden and Spruch~\cite{schaden1,schaden2} 
for the case of two spheres subject to Dirichlet boundary conditions 
in the semiclassical
framework of Gutzwiller's trace 
formula~\cite{gutbook}. It was shown that the result of the proximity force 
approximation emerged from the semiclassical calculation in the limit of
vanishing surface-to-surface separation. Gies and coworkers numerically
studied the
scalar sphere-plate case, again 
subject to Dirichlet boundary conditions, 
in the world-line approach~\cite{gies}. Scardicchio and Jaffe analyzed these
results in an optical approach that takes into account not only periodic, but
also closed orbits~\cite{scar}.

\begin{figure}[h,t,b]

\centerline{\includegraphics[width=8.0cm]{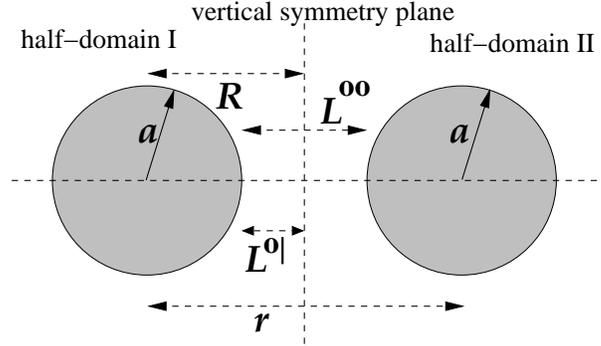}}

\caption{Two spheres of common radius $a$ at a center-to-center separation $r$.
The vertical symmetry plane splits the full domain into
two half-domains I and II. The 
center-to-plane separation $R$,
the surface-to-surface separation  $L^{00}$ in
the full domain and  $L^{0|}$ in the half-domain are indicated as well.} 
\label{fig:two-spheres}
\end{figure}

In the case of two spheres there is a continuous axial symmetry with respect to
the line joining the centers of the spheres, and an additional 
reflection symmetry with respect to any plane containing this (horizontal)
symmetry axis. The symmetry group is
$C_{\infty h}$ in
crystallography group theory notation~\cite{hamermesh}, where the role
of the vertical $v$ plane and horizontal $h$ plane 
is opposite to the one of this reference 
in order to match the orientation of figure~\ref{fig:two-spheres}. 
As a consequence of the $C_{\infty h}$ symmetry the
inverse multi-scattering matrix is separable with respect to the
magnetic 
quantum number $m$:
\begin{equation}
\left(\begin{array}{cc}M^{11} & M^{12}\\
                       M^{21} & M^{22}\end{array}\right)_{lm,l'm'}
= \delta_{m m'} \left(\begin{array}{cc}\delta_{ll'} & A^{12}_{lm,l'm}\\
                         A^{21}_{lm,l'm}& \delta_{ll'}\end{array}\right)
\end{equation} 
where the indices $1,2$ label the two spheres and where $M^{jj'}_{lm,l'm'}$
is the left hand side of (\ref{Msphere}), whereas 
$A^{jj'}_{lm,l'm'}$ is the non-trivial term (minus the delta-distribution 
term
$\delta^{jj'}\delta_{ll'}\delta_{mm'}$)
on the right-hand side of this equation -- both adjusted to 
the two-sphere case.

The case of two identical spheres at a center-to-center distance $r$ contains
the case of a sphere and a plate at a center-to-plate distance $R=r/2$ 
as can be seen in
figure~\ref{fig:two-spheres}. 
Because of the additional two-fold reflection symmetry with respect to
the vertical symmetry plane, the symmetry group of the two-sphere case with
common radius $a$ is even $D_{\infty v}$. Therefore, the global domain splits
into two half-domains (called I and II) that are 
separated by the vertical symmetry
plane. Moreover, 
all scattering waves can be separated into symmetric ones (subject
to Neumann boundary conditions)  and 
antisymmetric ones (subject to Dirichlet boundary conditions) 
with respect to this plane. Thus there exist two classes of  inverse 
multi-scattering matrices in one half-domain, one subject to Neumann boundary
conditions (N) and one subject to Dirichlet boundary conditions (D) on the 
vertical symmetry plan~\cite{BMW06}:
\begin{eqnarray}
  \left. M^{{\rm oo}\,(m)}_{ll'}\right|_{\rm N} &=& \delta_{ll'}
  + A_{ll'}^{(m)}\,,\quad
   \left. M^{{\rm oo}\,(m)}_{ll'}\right|_{\rm D}
   = \delta_{ll'} - A_{ll'}^{(m)}
\end{eqnarray}
(using the short-hand notation
$A_{ll'}^{(m)}\equiv A_{lm,l'm}^{12}=A_{lm,l'm}^{21}$),
and the associated determinants factorize as
\begin{eqnarray}
\det M^{{\rm oo}}(k,a,r) &=& \prod_{m=-\infty}^\infty 
  \det M^{{\rm oo}\,(m)}(k,a,r) \nonumber\\
&=& \det \left. M^{{\rm oo}}(k,a,r)\right|_{\rm N}
                     \det \left. M^{{\rm oo}}(k,a,r)\right|_{\rm D}\,.
\end{eqnarray}
Therefore the determinant for the sphere-plate case under Dirichlet boundary
conditions can be inferred from 
$\det \left. M^{{\rm oo}}(k,a,r)\right|_{\rm D}$
as follows:
\begin{equation}
\det M^{{\rm o|}}\Bigl(k,a,L^{\rm o|}\Bigr) =\left. \det M^{{\rm oo}}
  \Bigl(k,a,r\!=\!2(L^{\rm o|}\!+\!a)\Bigr)\right|_{\rm D}
\end{equation}
and the pertinent Casimir energy is given as \cite{BMW06}
\begin{equation}
{\cal E}_{\rm C}^{{\rm o |}}(a,L)
= \frac{\hbar c }{2\pi}
  \int_0^\infty \!\!\rmd k_4 \ln\det \left. M^{\rm o o} 
\Bigl(\rmi k_4,a,r\!=\!2(L\!+\!a)\Bigr)\right|_{\rm D} \,.
\label{ESsphplat}
\end{equation}

The exact data generated by this equation are reported in \cite{BMW06}.
They are compatible with the numerical data 
of the worldline
approach~\cite{gies} 
between $L=a/8$ and $L=a$ if the statistical error bars of the worldline data
are taken into account, see \cite{BMW06}. Moreover,  
the (systematically and statistically improved) worldline data of 
\cite{gies2005,gies2006}, 
which extend up to $L=16a$ and up to $L<100 a$, respectively,
do  nicely agree with the exact data of \cite{BMW06}.
In principle, the calculation based on
(\ref{ESsphplat}) is exact and applicable for all separations of
the obstacles. In practice, it is more suited for large- to medium-range
separations.
The reason is that the Casimir energy for the scalar case is dominated
by momenta $k\sim 1/L$ where $L$ is the separation scale. As shown in
\cite{BMW06} for the scalar
sphere-plate case, the Casimir integration can be 
truncated at $k_{\rm max}\sim 10/L$
corresponding to a truncation in the angular momentum space at 
$l_{\rm max} \leq (\rme/2) k_{\rm max} a\approx 14 a/L$.
For $L<0.1a$ the matrices increase rapidly in size and the angular-momentum
algebra (see the 3j symbols in (\ref{Msphere})) becomes very cumbersome. 

Let us rather concentrate on the asymptotical behaviour (surface-to-surface
separation $L\gg a$)
of Casimir energy given by (\ref{ESsphplat}).
The $s$-wave approximation
is not needed for the {\em fermionic Casimir energy}  since it is 
governed by the UV part of
the density of states ({\it i.e.} by the contribution at  the Fermi
momentum $k_{\rm F}$, assuming $k_{\rm F}>1/a$).
However, it dominates the asymptotics of  the fluctuating-scalar
Casimir effect at very large separations ($L\gg a$), as
this Casimir-energy type is governed by the infrared behaviour
of the density of states at $k\sim 1/L$ or less.
The Casimir energy of the
sphere-plate case subject to Dirichlet boundary conditions is given at
large separations $L\gg a$ (in the $s$-wave approximation) as~\cite{BMW06}
\begin{eqnarray}
{\cal E}(L) &\sim& -\frac{\pi^3\hbar c\, a}{1440 L^2}\,
{\frac{90}{\pi^4}\frac{2}{(1+a/L)(1+a/2L)}}\nonumber\\
     &\to&  -\frac{\pi^3 \hbar c\, a}{1440 L^2}\,{\frac{90}{\pi^4}
\times 2}
=  -\frac{\pi^3 \hbar c\, a}{1440 L^2} \times {1.847\cdots}\,.
\end{eqnarray}
As in section~\ref{sec:fermionic} there is a  
relative factor of 2 originally found in \cite{BW01}
between the $s$-wave
and semiclassical result for each reflection off a sphere.
This applies also for
the fluctuating-scalar Casimir effect, if the
relativistic dispersion $E=\hbar c k$ and the suppression of the repeats of
the semiclassical two-bounce orbit at large separations (by the removal
of the term $\sum_{n=1} n^{-4}=\pi^4/90$) are taken into 
account~\cite{BMW06}.
Therefore,
the exact result of the fluctuating-scalar 
Casimir energy for the {\em very far separated} Dirichlet sphere-plate
system is enhanced by
a factor of $2$ relative to the semiclassical result  and
by a factor of
$2\times (90/\pi^4)$   relative 
to the {\em leading} term $-\pi^3\hbar c a/(1440 L^2)$ 
of the proximity-force 
approximation (PFA)~\cite{proximity}. The leading term of the PFA,
in other words the PFA 
at {\em vanishing distances}, where the repeats of the two-bounce
orbit cannot be neglected and add up to the  factor  $\pi^4/90$,
was -- as mentioned -- confirmed semiclassically in \cite{schaden1}.
In this connection, 
it should be noted  that the result
quoted in \cite{BMW06} for the semiclassical approximation to
scalar Dirichlet sphere-plate case
\begin{equation}
  {\cal E}_{\rm sc}^{{\rm o} |}(a,L) =
  - \frac{\hbar c}{16\pi}\,\frac{a}{L^2}
  \left(\frac{\pi^4}{90}\right)
  \left[1
  -\left(\frac{5}{\pi^2}-\frac{1}{3}\right)\,\frac{L}{a}
  +{\cal O}\left( [L/a]^2\right)
  \right]
  \label{E_sp_sc}
\end{equation}
also applies for the scalar Neumann case, since both semiclassical 
calculations
differ only by the Maslov indices ({\it i.e} by a minus sign 
for each reflection
from a Dirichlet surface) and since the number of reflections
is even for a self-retracing orbit. The electromagnetic case in the
semiclassical approximation
is then given by the sum of both scalar cases. It is therefore twice as big as
(\ref{E_sp_sc}) and the first correction to the leading PFA result is
predicted to be negative.

In the case of the electromagnetic Casimir effect for the sphere-plate 
system, the
$s$-wave dominance at large separation has to be replaced by a
$p$-wave dominance, since the charge-neutrality of the sphere forbids a
monopole
term, whereas the standard Casimir-Polder energy is  dominated by 
induced-dipole contributions. If one removes by hand the $s$-wave contribution
from the Casimir energy of the scalar Dirichlet sphere-plate system, the
remaining energy is dominated at large separations by the $p$-wave contribution
\begin{equation}
{\cal E}_{\mbox{\tiny p-wave}}(L)\sim -\frac{5\pi^3 \hbar
    c\, a^{3}}{1440 {L^4}}\,{\frac{90}{\pi^4}}\,.
\end{equation}
Note that this expression is compatible with the $a^3/L^4$ scaling of the
Casimir-Polder energy for a molecule-plate system~\cite{polder}, 
but the prefactor is different~\cite{datta}.

Together with M~Bordag~\cite{bordagpriv} the asymptotics of the Dirichlet
sphere-plate problem was worked out to sixth subleading order
in 
the ratio of the sphere 
radius $a$ and the center-to-plate separation $R=r/2$:  
\begin{eqnarray}
 E_{{\rm D}, l{\geq} 0} &=& - \frac{\hbar c a}{8 \pi R^2}
        \biggl\{ 1 + \frac{5}{8} \frac{a}{R}
                  + \frac{421}{144} \left(\frac{a}{R}\right)^2
                  + \frac{535}{1152} \left(\frac{a}{R}\right)^3
                  + \frac{3083041}{518400} \left(\frac{a}{R}\right)^4
\nonumber \\[1mm]
   &&\mbox{}\qquad
                  - \frac{2741117}{1382400} \left(\frac{a}{R}\right)^5
                  + \frac{557222415727}{36578304000}\left(\frac{a}{R}\right)^6
                  + \cdots\biggr\}\,.
 \label{ED_asymp}
\end{eqnarray}
Even 
higher-order contributions (up to ninth order) 
were considered in this asymptotic series, 
but did not lead to any 
improvements.
\begin{figure}[h,t,b]

\psfrag{L/a}{$L/a$}
\psfrag{E_C\(a,L\)/\(-hbar c pi^3 a/1440 L^2\)}
{${\cal E}_{\rm C}^{\rm o |}(a,L)\, /\,(- \frac{\hbar c\pi^3 a}{1440 L^2})$}

\centerline{\includegraphics[width=8.0cm,angle=-90]{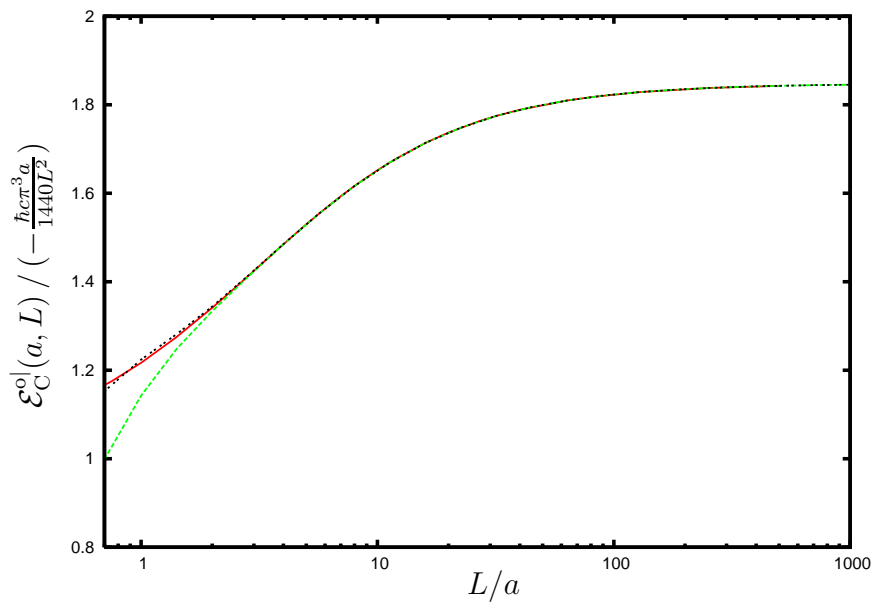}}

\caption{Exact and asymptotic results for the scalar Casimir energy of the 
sphere-plate configuration with Dirichlet boundary conditions
are shown shown in units of $-\hbar c \pi^3 a/(1440 L^2)$
as function of the ratio of the surface-to-surface-separation $L$ and the 
sphere radius $a$. The exact result from (\ref{ESsphplat}) 
is plotted as full line 
(in red),
the asymptotic approximation (\ref{ED_asymp}) up to the $(a/R)^4$ 
term 
as dashed line
(in green) and up to the $(a/R)^6$ term as short-dashed line (in blue).}
\label{fig:asymp}
\end{figure}
For practical purposes the sixth order expression is still 
useful down to separations $L=0.7a$ (or $R=1.7a$), see figure~\ref{fig:asymp}.
If the $s$-wave contribution is removed by hand, the corresponding 
expansion reads~\cite{bordagpriv}
\begin{eqnarray}
 E_{{\rm D}, l> 0} &=& - \frac{5 a^3}{16 \pi R^4}
        \biggl\{ 1 + \frac{56}{25} \left(\frac{a}{R}\right)^2
                  - \frac{597}{640} \left(\frac{a}{R}\right)^3
              + \frac{10453}{1750}\left(\frac{a}{R}\right)^4\nonumber\\[1mm]
  &&\mbox{}\qquad         - \frac{16557}{1600}  \left(\frac{a}{R}\right)^5
                  + \frac{394844679647}{9144576000}\left(\frac{a}{R}\right)^6
                  + \cdots\biggr\}\,.
\end{eqnarray}
Finally, for the corresponding Neumann sphere-plate system the removal of
the $s$-wave contribution is natural and the corresponding result is given
by~\cite{bordagpriv}
\begin{eqnarray}
 E_{{\rm N}, l > 0} &=& - \frac{10 a^3}{16 \pi R^4}
        \biggl\{ 1 + \frac{63}{100} \left(\frac{a}{R}\right)^2
                  + \frac{597}{320} \left(\frac{a}{R}\right)^3
                  - \frac{4159}{14000} \left(\frac{a}{R}\right)^4\nonumber\\
  &&\mbox{}\qquad  - \frac{271437}{25600}\left(\frac{a}{R}\right)^5
                  + \frac{148355331834}{2286144000} \left(\frac{a}{R}\right)^6
                  + \cdots\biggr\}\,.
\end{eqnarray} 
For reasons of completeness, we also report the numerical 
result of the pure $s$-wave
contribution for the sphere-plate case under Neumann boundary conditions, 
although it is of course artificial:
\begin{equation}
 E_{{\rm N}, l = 0}
= - \frac{1}{4 \pi R} \times 0.46066\ldots\times \biggl\{ 1
+ {\cal O}\left(\frac{a}{R}\right)\biggr\} \, .
\end{equation} 
Note that in this case the infrared limit $k_4\to 0$ and the asymptotic 
limit $R/a\to \infty$ do not commute in (\ref{ESsphplat}).

Finally, for comparison, the result of the Casimir-Polder calculation
in the limit of a perfectly conducting sphere as calculated 
in \cite{datta}  is listed here as well:
\begin{equation}
E_{{\rm EM},l>0}  =  - \frac{(3+6) a^3}{16 \pi R^4}\times \biggl\{ 1
+ {\cal O}\left(\frac{a}{R}\right)\biggr\}\,.
\end{equation}
We see that the relative ratio of 
the leading Dirichlet and Neumann $l>0$ contributions is correct, 
but that the scalar
result is bigger than the electromagnetic one by a  factor of $5/3$.

\section{Final remarks}
\label{sec:fin_rem}

In the meantime, after the work \cite{BMW06} appeared,
further exact Casimir results for non-separable systems have been reported.
In \cite{emig1}, the Casimir interaction between a plate and
a cylinder was calculated for the background of a fluctuating
electromagnetic field. The exact zero-point interaction between two
non-concentric cylinders was studied in \cite{mazzitelli}.
The Casimir effect for a sphere and a cylinder in front of a plane
and corrections to the proximity force theorem were investigated in
\cite{bordagnew}. The Casimir forces between arbitrary compact
objects were investigated in \cite{emig2}; see also the 
study~\cite{kenneth} in the $T$ operator approach and \cite{emig3,milton}.

Finally, let me mention the way in which the Casimir energy per unit length 
for $n$ non-overlapping
parallel cylinders of infinite length  
can be derived from the modified Krein formula of the 
two-dimensional $n$-disk case. 
For a cut perpendicular to the cylinder axes
the modified Krein formula of the $n$-disk case 
can be applied, provided the wave number $k$ in
that  formula is 
replaced by the modulus of the perpendicular wave number $k_\perp$ (the
wave number along the cut). Moreover, the phase space integration 
$L_\parallel\int \rmd k_\parallel/(2\pi)$ along
the parallel direction to the cylinder axes has to be added and the resulting
integrals have to be weighted by the
correct energy-dispersion $\hbar c \sqrt{k^2_\parallel + k^2_\perp}$. 
The corresponding Casimir energy per unit length therefore reads
\begin{eqnarray}
{\cal E}_{\rm C}/L_\parallel &=&  \int_{-\infty}^\infty 
\frac{\rmd  k_\parallel}{2\pi} \int_0^\infty
\rmd k_\perp \half \hbar c \sqrt{k_\parallel^2+k_\perp^2}\delta
\rho_{\rm C}(k_\perp,n\mbox{-disk})\nonumber\\
&=& \int_{-\infty}^\infty 
\frac{\rmd k_\parallel}{2\pi} \int_0^\infty
\rmd k_\perp \half \hbar c \sqrt{k_\parallel^2+k_\perp^2} \frac{d}{d k_\perp}
\frac{-1}{\pi} {\rm Im} \ln 
\det M(k_\perp,n\mbox{-disk})
\nonumber \\
&=&\frac{\hbar c}{2\pi^2}\int_{-\infty}^\infty
\rmd k_\parallel \int_{|k_\parallel|}^\infty \rmd k_4\,
\frac{k_4}{\sqrt{k_4^2-k_\parallel^2}}\ln \det M(\rmi k_4,n\mbox{-disk})
\nonumber \\
&=&\frac{\hbar c}{4\pi} \int_0^\infty \rmd k_4 k_4 \ln 
\det M(\rmi k_4,n\mbox{-disk})\,.
\end{eqnarray}

\section{Summary}
We have shown that the Casimir energy can be re-defined as the vacuum
energy of the geometry-dependent part of the density of states and that
the latter can be calculated form the multi-scattering phase shift of
a modified Krein formula.
The non-overlapping, in general non-separable $n$-sphere, sphere-plate, 
$n$-disk (and $n$-cylinder) Casimir problems can be solved exactly in
the scalar (and also in the fermionic) case.
The calculation is not plagued by the subtraction of single-sphere 
contributions or by the removal of diverging ultraviolet contributions.
All involved determinants exist and are finite since the pertinent T-matrices
are trace-class. The large-distance behaviour is dominated by the $s$-wave
scattering  in the case of the scalar Casimir effect with Dirichlet
boundary conditions  and by the
$p$-scattering for the corresponding Neumann case as in the electromagnetic
scenario.
The Dirichlet or Neumann boundary conditions can even be replaced by
mixed boundary conditions.
The presented method can easily be applied to any number of spheres
or cylinders with or without planes (in two dimensions: disks with or without
lines).
Moreover, 
the spheres (or disks) can be replaced by other objects or even smooth
potentials or non-ideal reflector, as long as these objects do not overlap.
The finite surface thicknesses can be booked as Weyl-term contributions.

\ack
{
A.W. would like to thank the organizer 
Michael Bordag for the invitation to this excellent workshop QFEXT'07
in Leipzig.}


\section*{References}

\begin{thebibliography}{10}
\bibitem{casimir48} 
    Casimir H B G 1948  {{\it Proc. Kon. Ned. Akad. Wetensch.}} 
{\bf 51} 793

\bibitem{bordag}
    Bordag M, Mohideen U  and Mostepanenko V M 2001 
       {\em Phys.\ Rept.} {\bf 353} 1

\bibitem{proximity}
    Derjaguin B V, Abrikosova I I and Lifshitz E M 1956 
       {{\em Quart. Rev.}} {\bf 10} 295;
    Blocki J, Randrup J, Swiateck W J and Tsang C F 1977 \APNY {\bf 105} 427 

\bibitem{BW01} 
    Bulgac A and Wirzba A 2001 
     \PRL {\bf 87} 120404
     [arXiv:nucl-th/0102018]

\bibitem{BMW05} 
 Bulgac A, Magierski P and Wirzba A 2005
 {\em Europhys. Lett.} {\bf 72} 327 
  [arXiv:cond-mat/0406255]


\bibitem{BMW06} 
    Bulgac A, Magierski P and Wirzba A 2006  \PR D {\bf 73}  025007 
    [arXiv:hep-th/0511056]

\bibitem{WBM06}
  Wirzba A,  Bulgac A and  Magierski P 2006
 \JPA {\bf 39}  6815
  [arXiv:quant-ph/0511057]


\bibitem{krein} 
    Krein M G 1953 {\em Mat. Sborn. (N.S.)\/} {\bf 33} 597;
    Krein M G 1962  {\em Sov. Math.-Dokl.\/} {\bf 3} 707; 
    Birman M Sh and Krein M G 1962 {\em Sov. Math.-Dokl.\/} {\bf 3} 740

\bibitem{uhlenbeck} 
    Beth E  and Uhlenbeck G E 1937 {\em Physica} {\bf 4} 915;
    Huang K 1987 {\it Statistical Mechanics}, 
       John Wiley \& Sons, New York, ch. 10.3; 
    Friedel J 1958 {\em Nuovo Cim. Ser. 10 Suppl.} {\bf 7}  287

\bibitem{gaspard}  
    Gaspard P and Rice S A 1989 \JCP {\bf 90} 2225; {\bf 90} 2242; 
                                     {\bf 90} 2255
\bibitem{wreport}  
    Wirzba A 1999 {\em Phys. Rept.} {\bf 309} 1
    [arXiv:chao-dyn/9712015]

\bibitem{wh98} 
    Wirzba A and Henseler M 1998 \JPA {\bf 31} 2155 
    [arXiv:chao-dyn/9702004]

\bibitem{hwg97} 
    Henseler M, Wirzba A and Guhr T 1997 \APNY {\bf 258} 286 
    [arXiv:chao-dyn/9701018] 

\bibitem{eckhardt}
    Eckhardt B 1987 \JPA {\bf 20} 5971 

\bibitem{cvitanovic}
    Cvitanovi\'c P and Eckhardt B 1989 \PRL {\bf 63} 823 

\bibitem{W92} Wirzba A 1992 {\em CHAOS} {\bf  2} 77

\bibitem{W93} Wirzba A 1993 \NP A {\bf 560} 136

\bibitem{gutbook} 
    Gutzwiller M C 1990 
       {\em Chaos in Classical and Quantum Mechanics} (Springer, New York)


\bibitem{schaden1}  
    Schaden M and Spruch L 1998 \PR A {\bf 58} 935


\bibitem{schaden2}
  Schaden M and Spruch L 2000
   \PRL {\bf 84} 459 

\bibitem{gies} 
    Gies H, Langfeld K  and Moyaerts L 2003 
    {\em J. High Energy Phys.} {\bf 06} 018 
    [arXiv:hep-th/0303264]

\bibitem{scar}
    Jaffe R L  and Scardicchio A 2004 \PRL  {\bf 92} 070402;
    Scardicchio A and Jaffe R L 2005 \NP B {\bf 704} 552 

\bibitem{hamermesh}
 Hamermesh M 1962, {\em Group Theory and Its Applications to Physical Problems}
(Addison-Wesley, Reading, MA), Chap. 9-1


\bibitem{gies2005}
     Gies H and Klingm\"uller K 2006 
     \JPA {\bf 39} 6415  
     [arXiv:hep-th/0511092]

\bibitem{gies2006}
     Gies H and Klingm\"uller K 2006 
   \PR D {\bf 74} 045002
    [arXiv:quant-ph/0605141]

\bibitem{polder}
  Casimir H B G and Polder D 1948
   \PR {\bf 73} 360 

\bibitem{datta}
Datta T and Ford L H 1981
\PL {\bf 83A} 314

\bibitem{bordagpriv}
Bordag M and Wirzba A unpublished

\bibitem{emig1}
Emig T, Jaffe R L, Kardar M and Scardicchio A 2006
  \PRL {\bf 96} 080403
[arXiv:cond-mat/0601055]

\bibitem{mazzitelli}
Mazzitelli F D, Dalvit D A R and  Lombardo F C 2006 
{\em New J.\ of Phys.} {\bf 8} 240
[arXiv:quant-ph/0610181]

\bibitem{bordagnew}
Bordag M 2006 
\PR  D {\bf 73} 125018
[arXiv:hep-th/0602295]

\bibitem{emig2}
Emig T, Graham N, Jaffe R L and  Kardar M 2007 
arXiv:0707.1862 [cond-mat.stat-mech] and
arXiv:0710.3084 [cond-mat.stat-mech]

\bibitem{kenneth}
Kenneth O and  Klich I 2007 
arXiv:0707.4017 [quant-ph]

\bibitem{emig3}
Emig T and Jaffe R L 2007
arXiv:0710.5104 [quant-ph]

\bibitem{milton}
Milton K A  and Wagner J 2007
  arXiv:0711.0774 [hep-th]

\end{thebibliography}
\end{document}